\documentclass[amssymb,amsfonts,12pt,a4paper]{article}
\usepackage{amsmath}
\usepackage{amsfonts}
\usepackage{amssymb}
\usepackage{array,dcolumn,psfrag}
\usepackage{graphicx}
\usepackage{booktabs,lscape}
\usepackage{algorithmic}
\usepackage{epsfig,amssymb,color}
\usepackage{multirow}
\include{setup}
\def\baselinestretch{1.5}

\def\lsp#1{\def\baselinestretch{#1}\@normalsize}

\title{Estimate  the Occurrence Rate of the DNA Palindromes }
\date{}
\author{
    I-Ping Tu\footnote{Corresponding author.
\textit{Email address}: iping@stat.sinica.edu.tw }, Yuan-Fu Huang and Shao-Hsuan Wang\\
    Institute of Statistical Science, Academia Sinica, Taipei, Taiwan\\
}

\begin{document}
\maketitle
\begin{abstract}
A DNA palindrome is a segment of  double-stranded DNA sequence with
inversion symmetry which may form secondary structures conferring
significant biological functions ranging from RNA transcription to
DNA replication. To test if the clusters of DNA palindromes
distribute randomly is an interesting bioinformatic problem, where
the occurrence rate of the DNA palindromes is a key estimator for
setting up a test. The most commonly used statistics for estimating
the occurrence rate for scan statistics is the average rate.
However, in our simulation, the average rate may double the null
occurrence rate of DNA palindromes due to hot spot regions of 3000
bp's in a herpes virus genome. Here, we propose a formula to
estimate the occurrence rate through an analytic derivation under a
Markov assumption on DNA sequence. Our simulation study shows that
the performance of this method has improved the accuracy and
robustness against hot spots, as compared to the commonly used
average rate. In addition, we derived analytical formula for the
moment-generating functions of various statistics under a Markov
model, enabling further calculations of p-values.

\end{abstract}
\bigskip


\section{Introduction}
A chromosome is a long sequence of double  helix  DNA made of base
pairing by an adenine-thymine($A=T$) pair or a
cytosine-guanine($C\equiv G$). Thus, one DNA strand decides the
sequence of its complementary strand. A segment of DNA sequence with
half length greater than or equal to a pre-specified length $L$ is
called a palindrome if one strand is identical to its complementary
one running at the reverse direction. It has been observed that DNA
palindromes are common candidates for searching genetic motifs
involved in different cellular processes, including gene
transcriptions, gene replications, and gene deletions. For example,
among nine octameres suggested to be transcription factor binding
sites, three are palindromes (FitzGerald et al, 2004). This might be
contributed by its potential to create the secondary genomic
structure (Leach, 1994).


Many studies have focused on investigating the occurrence rates of
palindromes in suspicious regions against random sequences. For
example, Lisnic and Svetec (2005) investigated the frequencies of
Palindromes in the yeast {\it Saccharmyces cerevisiae} genome
according to the length and contents of palindromes. Chew et al
(2005) proposed three score schemes, based on occurrence rates,
length or its likelihood, to quantify the palindromes and found the
association between the high score regions and the replication
origins. Lu et al (2007) reported that meaningful sites tend to have
higher palindrome scores by comparing the scores over the regions
including introns, exons, and upstream of transcription start sites
against simulated random sequences.

The performance of these comparison tests strongly depends on how
accurate the occurrence rate is estimated for the random sequence.
This rate is usually estimated by the average rate of palindromes on
the genome-wide sequence. Another approach is the iid model based
estimator which a formula has been derived when the DNA letter
frequencies are estimated (Chew, et al, 2005).  However, we observed
obvious discrepancies between these two estimates in various herpes
virus genomes. For an example on the BHV1CGEN(BoHV1) sequence,
average rate is 0.00166 and the iid model method estimate  the rate
as 0.00073. While the average rate might be bias due to hot spot
regions, the iid model might be too naive to describe the DNA
sequence. In this paper, we {provided} a formula to calculate the
occurrence rate under a Markov model, which the iid model would
become a special case.  For the BoHV1 case, our method estimates the
rate as 0.00098. Simulations are designed to check the performance
of the estimates on the null occurrence rate, including with and
without hot spot segments in the random sequences. The results show
 that our method performs better than the average rate in estimating the
 null occurrence rate against hot spot regions.

Chan and Zhang (2007) developed a method to approximate the p-value
of statistics for weighted Poisson process, which can be applied on
the DNA palindrome problems. In their approach, the analytic formula
for the moment generating function (MGF) of the palindrome score is
required. However, the distribution of the palindrome scores have
not been well studied except the length score under iid assumption.
Thus, we developed a method to derive the analytic formula for the
MGF on various scores under Markov model.
 Furthermore, this
analytic formula allows us to calculate an overshoot term in the
p-value approximation.

This paper is organized as follows.  In section 2, we show that
three commonly used scores proposed by Chew et. al. (2005) can be
derived by a likelihood approach firstly. Secondly, we show that the
occurrence rates can be calculated accurately under Markov model
through constructing a quasi transition matrix $T$. Thirdly, we
derive the moment generating function for various scores under the
Markov model. Last, we gave a p-value approximation with more
precise calculations on the overshoot term. In section 3, we show
the numerical study for both real data and simulated data. This
paper ends with a brief discussion.

\section{Method}

\subsection{Notations and Log Likelihood Ratio Statistics}

Let $N(t)$ be a counting process to describe the occurrence of
palindromes and let $N_w(t)=N(t+w)-N(t)$ denote the number of events
in the interval $(t, t+w]$. Leung et al (2005) proved that $N(t)$
can be approximated by a Poisson process under Markov Model. We let
$x_i$ be the score for the $i^{th}$ event along the genome sequence.
$S_{N_w(t)}$ is the summation of the Palindrome scores inside the
interval $(t, t+w]$, which can be expressed by equation (\ref{101}):
\begin{eqnarray}
S_{N_w(t)}=\sum_{i=N(t)+1}^{N(t+w)}x_i.\label{101}
\end{eqnarray}

To search the clusters of palindromes, Chew et al (2005) proposed 3
schemes on scoring palindromes for prediction of replication origins
in herpes viruses. They are palindrome count score(PCS), palindrome
length score(PLS), and base-pair weighted score of order m
($BWS_m$). PCS gives score one for each DNA palindrome; PLS gives
the score as the palindrome length divided by its minimum required
lenth; whereas $BWS_m$ gives the score as the minus log-likelihood
with Markov order $m$.

We would like to show that both $N_w(t)$ and $S_{N_w(t)}$ are
equivalent to some log-likelihood ratio statistics when the
alternative hypotheses are properly constructed. Under the Poisson
process model, $x_i$'s can be treated as iid with a density function
{$f_{\theta}(x)=f_0(x)\exp(\theta x - \phi(\theta))$, where $f_0(x)$
is an unknown distribution and $\phi(\theta)=\log \int e^{\theta
x}f_0(x)dx$.} The parameters for $N(t)$ and $x_i$ are $(\lambda_a,
\theta_a)$ for those events occurred in the interval $(t_a, t_a+w]$
and $(\lambda_0, \theta_0)$ otherwise; and the null hypothesis is
$\lambda_a=\lambda_0$ and $\theta_a=\theta_0$. When $t_a$ is known,
the likelihood ratio is
$f_{\lambda_a,\theta_a}(N_w(t_a),S_{N_w(t_a)})/
f_{\lambda_0,\theta_0}(N_w(t_a),S_{N_w(t_a)})$, where the likelihood
is as follows:
\begin{eqnarray}
&&f_{\lambda,\theta}(N_w(t),S_{N_w(t)})\nonumber\\
&=&f_{\lambda}(N_w(t))
f_{\theta}(S_{N_w(t)}|N_w(t))\nonumber\\
&=&{(\lambda w)^{N_w(t)}e^{-\lambda w}\over N_w(t)!}{\{\prod
f_0(x_i)\}}\exp(\theta
S_{N_w(t)}-N_w(t)\phi(\theta)).\nonumber\end{eqnarray}

Because $t_a$ is usually unknown, we search the maximum of the
statistic over all possible $t$.
\begin{itemize}
\item [Case 1.]
If the alternative hypothesis is constructed as $H_a:~$
$\lambda_a=\lambda_1>\lambda_0$ and $\theta_a=\theta_0$, then the
log-likelihood ratio statistic is equivalent to PCS in Chew et al
(2005), which is shown as follows
\begin{equation}\label{103}
\max_{t}l_t(\lambda_1,\theta_0)=\max_t\log\left({f_{\lambda_1,\theta_0}(N_w(t),S_{N_w(t)})
\over f_{\lambda_0,\theta_0}(N_w(t),S_{N_w(t)})}\right)=\max_t
N_w(t)\log({\lambda_1\over \lambda_0})-(\lambda_1-\lambda_0){w}.
\end{equation}

\item [Case 2.]
If the alternative hypothesis is constructed as $H_a:~$
$\lambda_a=\lambda_1>\lambda_0$ and $\theta_a=\theta_1>\theta_0$,
where $\lambda_1$ and $\theta_1$ are with the constraint
\begin{equation}\label{constraint}
\log({\lambda_1\over\lambda_0})- (\phi(\theta_1)-\phi(\theta_0))=0,
\end{equation}
the log-likelihood ratio statistic in formula (\ref{104}) can be
equivalent to PLS or $BWS_m$ proposed by Chew et al (2005),
depending on the definition of $x_i$'s.
\begin{eqnarray}\label{104}
\max_tl_t(\lambda_1,\theta_1)&=&\max_t\log\left({f_{\lambda_1,\theta_1}(N_w(t),S_{N_w(t)})
\over
f_{\lambda_0,\theta_0}(N_w(t),S_{N_w(t)})}\right)\nonumber\\
&=& {\max_t
\big\{-(\lambda_1-\lambda_0)w+(\theta_1-\theta_0)S_{N_w(t)} \big\}}
\end{eqnarray}
\end{itemize}

It can be observed that (\ref{103}) is equivalent to $\max_t N_w(t)$
and (\ref{104}) is equivalent to $\max_tS_{N_w(t)}$. While
(\ref{103}) only tests the Poisson parameter $\lambda$, (\ref{104})
tests both the Poisson parameter $\lambda$ and score parameter
$\theta$ with the constraint (\ref{constraint}). It may be helpful
to be reminded that $ N_w(t)$ can be treated as a special case of
$S_{N_w(t)}$ with $x_i=1$ for each $i$.

Chan and Zhang (2007) developed an approximation method to calculate
p-value of the scan statistics on a weighted Poisson process, which
can be applied to derive the threshold value of (\ref{101}) if the
MGF $\phi(\theta)$ of $x_i$ is properly formulated. Let $N(t)$ be a
Poisson process with mean $\lambda_0$ and moment generating function
(MGF) $x_i$'s are iid with {mean $\mu_0$}, then

\begin{eqnarray}\label{102}
&&\mathrm{P}_0(\max\limits_{0 <t< W}S_{N_w(t)} \geq b)\nonumber\\
&\sim& 1-exp\left(-(W-w) \nu_{\lambda_1,\theta_1}( b- \lambda_0
\mu_0)e^{-[b\theta_1-w(\lambda_1-\lambda_0)]} (2 \pi w \lambda_1
\phi''(\theta_1))^{-1/2}\right),
\end{eqnarray}
where $W$ is the total length of the sequence {and}
$\nu_{\lambda_1,\theta_1}$ is an overshoot correction term and
$\theta_1$ and $\lambda_1$ satisfy the equations:
\begin{eqnarray*}
&&\lambda_1\phi'(\theta_1)=b,\\
&&\log(\lambda_1/\lambda_0)={\phi(\theta_1)-\phi(\theta_0)}.
\end{eqnarray*}

Whether $ N_w(t)$ or $S_{N_w(t)}$ is used in testing the null
hypothesis, $\lambda_0$ always plays a crucial role.  If $\lambda_0$
is overestimated seriously, the test would be too conservative and
lose its power. Alternatively, if $\lambda_0$ is underestimated
seriously, the test would fail.

\subsection{Occurrence rate of DNA palindromes under Markov model}

The average rate is a commonly used estimator for the null parameter
of scan statistics. Yet, in various herpes virus genomes, it can be
observed that the average rate is positive bias affected by some hot
spot regions. On the other hand, the iid mode may not be a good
model to describe the DNA sequence well since it ignores the
correlation between adjacent DNA letters. Thus, we developed a
method to calculate the occurrence rate of the palindromes under a
Markov model. We constructed a matrix $T$, with
$T_{ij}=P_{a_ia_j}P_{\tilde{a}_j\tilde{a}_i}$ which groups together
the transition probabilities of symmetric complimentary pairs. For
example, AG  would conjugate with CT on its mirror site which leads
to define $T_{13}=P_{AG}P_{CT}$, and we call $T$ a quasi transition
matrix because its row does not sum to one.

\noindent{\bf Theorem 1} Assume that DNA letters along the genome
sequence follow a Markov model with transition probability $\{P_{a,
b}|a,b \in \{A,C,G,T\}\}$ and the letter frequency $P_0'=
(\pi_{A}~\pi_{C}~\pi_{G}~\pi_{T} )$, then the occurrence probability
of a palindrome given a starting position with half length greater
or equal to $L$ is
 \begin{equation}\label{201}
 \lambda_M\equiv P\left(\parallel I \parallel \ge L \right)
 = P'_0 T^{L-1} P_1
 \end{equation} {where $I$ describes the palindromic pattern given a starting position and $\| I\|$ denotes
 the corresponding maximum length},
 \begin{eqnarray}
P_1'&=& \left(P_{AT}~P_{CG}~P_{GC}~P_{TA} \right),\nonumber
\end{eqnarray} and
$$ T = \left(\begin{array}{cccc}
P_{AA}P_{TT} & P_{AC}P_{GT} & P_{AG}P_{CT} & P_{AT}P_{AT} \\
P_{CA}P_{TG} & P_{CC}P_{GG} & P_{CG}P_{CG} & P_{CT}P_{AG} \\
P_{GA}P_{TC} & P_{GC}P_{GC} & P_{GG}P_{CC} & P_{GT}P_{AC} \\
P_{TA}P_{TA} & P_{TC}P_{GA} & P_{TG}P_{CA} & P_{TT}P_{AA}
\\ \end{array} \right).$$

\noindent{\bf Proof:}  The set that a DNA palindrome with half
length greater or equal to $L$, is equivalent to the set that the
center $2L$ letters follows a palindrome pattern. Given a sequence
of length $2L$, it must satisfy that $a_{L+k}={\tilde a}_{L-k+1}$ to
become a palindrome, $\tilde{a}_{i}$ means the complementary letter
of $a_i$. Then, under a Markov model, we can sum the probability
over all possible the letters and get $\lambda_M$.

\begin{eqnarray}
\lambda_M=&& P\left(\parallel I \parallel \ge L
\right)\nonumber\\
&=&\sum_{\scriptstyle a_i\in{\{A,C,G,T\}} \atop \scriptstyle 1\le i
\le L}\pi_{a_{1}}P_{a_{1}a_{2}} \ldots
P_{a_{L-1}a_{L}}P_{a_{L}\tilde{a}_{L}}P_{\tilde{a}_{L}\tilde{a}_{L-1}}
P_{\tilde{a}_{L-1}\tilde{a}_{L-2}} \ldots
P_{\tilde{a}_{2}\tilde{a}_{1}}\nonumber\\
&=&\sum_{\scriptstyle a_i\in{\{A,C,G,T\}} \atop \scriptstyle 1\le i
\le L}\pi_{a_{1}}\left(P_{a_{1}a_{2}}
P_{\tilde{a}_{2}\tilde{a}_{1}}\right)\ldots
\left(P_{a_{L-1}a_{L}}P_{\tilde{a}_{L}\tilde{a}_{L-1}}\right)
P_{a_{L}\tilde{a}_{L}} \label{202}\\
 &=& P'_0 T^{L-1} P_1\nonumber
\end{eqnarray}

$P_{a_i,a_{i+1}}$ is the transition probability for letter $a_i$ to
letter $a_{i+1}$. $T$ is the matrix form of $(P_{a_{1}a_{2}}
P_{\tilde{a}_{2}\tilde{a}_{1}})$. $(\ref{202}$) can be viewed as a
matrix multiplication: a row vector multiplies a matrix to the power
of L and then multiplies with a column vector. This technique is
used repeatedly in this paper, including the proof for Theorem 3.

 \noindent{\bf Remark 1:} When the Markov model is
reduced to the iid model, $P_1'$ becomes
$$P_2'=\left(\begin{array}{cccc}
\pi_{T} & \pi_{G} & \pi_{C} & \pi_{A} \\
\end{array} \right),
$$ and $T$ becomes $P_2P_0'$. Thus,
{
\begin{equation}\label{203}
 \lambda_{\text{iid}}\equiv P\left(\parallel I \parallel
\ge L
\right)=P_0'(P_2P_0')^{L-1}P_2=(P_0'P_2)^L=\gamma^{L},
\end{equation}
} where $ \gamma = 2\left(\pi_{A}\pi_{T}+\pi_{C}\pi_{G}\right)$.
$(\ref{203})$ has been shown in
Leung et al(2005).\\

\noindent{\bf Theorem 2} With the same assumption in Theorem 1, the
PLS score for the $i^{th}$ palindrome is defined as $x_i=\|I_i\|/L$
conditional on $\|I_i\|\ge L$, where $L$ is the minimum half length
for the palindrome. Then, the MGF for $x_i$ is
\begin{equation}\label{204}
K_{PLS}(t)\equiv E\left(e^{x_it}|\|I_i\|\ge L\right)= {e^{t}\over
\lambda_M}P'_0T^{L-1}[I-e^{t/L}T]^{-1}[I-T]P_1
\end{equation}

\noindent{\bf Proof of Theorem 2}

\begin{eqnarray}
 &&E\left(e^{x_it}| \|I_i\|\geq L \right) \nonumber\\
 &=&\sum_{k=L}^{\infty}e^{kt/L}[P(\|I_i\|\ge k)-P(\|I_i\|\ge
 k+1)]/P(\|I_i\|\ge L)\nonumber\\
&=&P_0' \sum_{k=L}^{\infty}e^{kt/L}T^{k-1}(I-T)P_1/\lambda_M\nonumber\\
&=&{e^{t}\over \lambda_M}P'_0T^{L-1}[I-e^{t/L}T]^{-1}[I-T]P_1
\end{eqnarray}

 \noindent{\bf Remark 2:} When the Markov model is reduced to iid
model,
$$K_{PLS}(t)=\sum_{k=L}^{\infty}e^{kt/L}(\gamma^k-\gamma^{k+1})/\gamma^L
={e^t(1-\gamma)\over 1-e^{t/L}\gamma}.$$

\noindent{\bf Theorem 3} With the same assumption in Theorem 1, the
BWS score is defined as $x_i=-log( P(I_i))$ conditional on
$\|I_i\|\ge L$. Then, the MGF for $x_i$ is
\begin{eqnarray}\label{205}
K_{BWS}(t) \equiv \mathrm{E}[e^{x_it} | \|I_i\| \geq L] =
\frac{1}{\lambda_{M}}{\bf v'}(t)[I-Q(t)]^{-1}[Q(t)]^{L-1}{\bf u}(t),
\end{eqnarray}
where
 ${\bf v}(t)=(v_1(t)~v_2(t)~v_3(t)~v_4(t))'$ is defined as
${v}_i(t)=\big( [(I-T)P_0]_i\big)^{1-t}$; $Q(t)$ is defined as
$Q_{ij}(t) = (T_{ij})^{(1-t)}$; and ${\bf
u}(t)=(u_1(t)~u_2(t)~u_3(t)~u_4(t))'$ is defined as ${ u}_i(t)
=\left([P_1]_i\right)^{1-t}$ with $i=1,\cdots, 4$.

 \noindent{\bf Proof of Theorem 3}

\begin{eqnarray*} &&P(I_i=a_1\cdots
 a_k\tilde{a}_k\cdots\tilde{a}_1,
 \|I_i\|=k)\\
 &=&\big\{(\pi_{a_1}-\sum\limits_{a_0 \in \{A,C,G,T\}}
 \pi_{a_0}\mathrm{P}_{a_0a_1}\mathrm{P}_{\tilde{a}_1\tilde{a}_0})
 \mathrm{P}_{a_1a_2}\cdots\mathrm{P}_{a_{k-1}a_k}\mathrm{P}_{a_k\tilde{a}_k}
 \mathrm{P}_{\tilde{a}_k\tilde{a}_{k-1}}
 \cdots\mathrm{P}_{\tilde{a}_2\tilde{a}_1}\big\}.
 \end{eqnarray*}

 Thus, we have
\begin{eqnarray}\label{206}
&&K(t,k)\nonumber\\
&\equiv& \mathrm{E}[e^{x_it} ; \|I_i\|=k]=\mathrm{E}[\left(P\{I_i \text{~occurs}\}\right)^{-t} ; \|I_i\|=k]\nonumber\\
&=& \sum\limits_{a_j \in \{A,C,G,T\} \atop {1\leq j \leq k}}
\big\{(\pi_{a_1}-\sum\limits_{a_0 \in \{A,C,G,T\}}
\pi_{a_0}\mathrm{P}_{a_0a_1}\mathrm{P}_{\tilde{a}_1\tilde{a}_0})
\mathrm{P}_{a_1a_2}\cdots\mathrm{P}_{a_{k-1}a_k}\mathrm{P}_{a_k\tilde{a}_k}\mathrm{P}_{\tilde{a}_k\tilde{a}_{k-1}}
\cdots\mathrm{P}_{\tilde{a}_2\tilde{a}_1}\big\}^{(1-t)}\nonumber\\
&=& \sum\limits_{a_j \in \{A,C,G,T\} \atop {1\leq j \leq k}}
\big(\pi_{a_1}-\sum\limits_{a_0 \in \{A,C,G,T\}}
\pi_{a_0}\mathrm{P}_{a_0a_1}\mathrm{P}_{\tilde{a}_1\tilde{a}_0}\big)^{(1-t)}
\left(\mathrm{P}_{a_1a_2}\mathrm{P}_{\tilde{a}_2\tilde{a}_1}\right)^{(1-t)}
\times \cdots \nonumber\\
&&~~~~~~~~~~~~~~~~\times \left(\mathrm{P}_{a_{k-1}a_k}\mathrm{P}_{\tilde{a}_k\tilde{a}_{k-1}}\right)^{(1-t)}
\left(\mathrm{P}_{a_k\tilde{a}_k}\right)^{(1-t)}\nonumber\\
&=& {\bf v'}(t) [Q(t)]^{k-1}{\bf u}(t).
\end{eqnarray}

Then, taking the sum over $k=L$ to $\infty$ and dividing by
$\lambda_M$ lead to $(\ref{205})$.

\noindent{\bf Remark 3}\\
When the Markov model is reduced to iid model, (\ref{206}) becomes
$$K(t,k)=(1-\gamma)^{1-t}P'_0(t)(P_2(t)P'_0(t))^{k-1}P_2(t)
=(1-\gamma)^{1-t}(P'_0(t)P_2(t))^k=(1-\gamma)^{1-t}\gamma_t^k,$$
where $P'_0(t)=(\pi_A^{1-t}~\pi_C^{1-t}~\pi_G^{1-t}~\pi_T^{1-t})$,
$P'_2(t)=(\pi_T^{1-t}~\pi_G^{1-t}~\pi_C^{1-t}~\pi_A^{1-t})$, and
$\gamma_t=P'_0(t)P_2(t)=2[(\pi_A\pi_T)^{1-t}+(\pi_C\pi_G)^{1-t}]$.
So, for iid model,
\begin{equation}\label{207}
K(t)={(1-\gamma)^{1-t}\over
1-\gamma_t}\left({\gamma_t\over\gamma}\right)^L.
\end{equation}

\noindent{\bf Remark 4}\\
The conditional process involved in the overshoot term in the
p-value approximation can be approximated by a partial sum of iid
copies of $y =  (-\sum\limits^{N(\Delta)}_{k=1}x_k +
\sum\limits^{N^*(\Delta)}_{k=1}x^*_k)$, where $N(\cdot)$ and
$N^*(\cdot)$ are iid Poisson processes with rates $\lambda_0$ and
$\lambda_1$; $x_i$'s and $x_i^*$'s are independent random variables
with density functions $f_{\theta_0}$ and $f_{\theta_1}$. The
derivation is in the appendix. By the same method in Theorem 3 and
Theorem 4, the characteristic function of $y$ can be derived.
Applying Theorem 1 in Tu(2009), the overshoot term can be
calculated.

\section{Real Data Analyses and Simulations}

We studied 27 herpesvirus genome sequences from the database of EBI
Nucleotide Sequences. For each sequence, we estimated the transition
matrix and the stationary probabilities of DNA letters
$\{\text{A,C,G,T}\}$. {\bf Theorem 1} is applied  to estimate the
null occurrence rate for each sequence. These results are compared
with those estimated by their average rates in Figure \ref{f1}. The
average rates show higher values consistently.

\begin{center}
\smallskip
\begin{figure}
\includegraphics[width=0.8\textwidth]{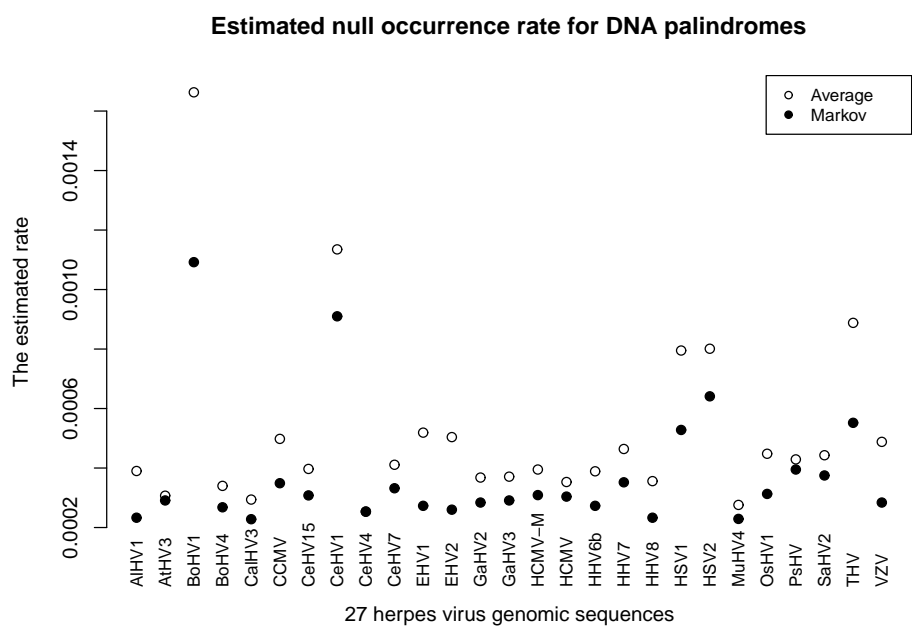}
\caption{ \label{f1} 27 herpes virus genomic sequences were
downloaded from {the database of EBI Nucleotide Sequences}. Two
methods for estimating the null palindrome rates are presented,
including the average rate, and the Markov model based estimator. We
adopted the abbreviation for naming the genome sequences used in
Leung et al. (2005)}
\end{figure}
\end{center}

We also checked the accuracy performance of these two methods
through numerical simulation. While a real DNA sequence may contain
meaningful DNA codes which contribute to its non-randomness, random
sequences are generated to fit the null hypothesis.  All the
parameters involved in generating the random sequences, including
the stationary probabilities $\pi$ and the transition matrix $P$,
are estimated on the BoHV1 sequence. BoHV1 sequence, with sequence
ID BHV1CGEN, contains 135301 bases. The state probabilities are
estimated as
  $$\pi= ( 0.1354(A) ,   0.3588(C),    0.3654(G),   0.1404(T))$$ and the transition
probabilities are
$$P=\left(\begin{array}{ccccc} &A&C&G&T \\
A & 0.1854& 0.3288& 0.3556& 0.1303\\
C & 0.1258& 0.2932& 0.4347& 0.1463\\
G & 0.1343& 0.4512& 0.2994& 0.1151\\
T & 0.1141& 0.3151& 0.3695& 0.2012\\
\end{array} \right).$$

 The half length $L=6$ is adopted
to be the criterion as a palindrome event.  Palindrome events along
these random sequences could be well approximated by a homogeneous
Poisson process. It may be helpful to be reminded that, in this
case, the average rate $\bar{\lambda}$ is the maximum likelihood
estimator (MLE) for the occurrence rate. Our simulation shows that
both these two methods do the estimate well in the first numerical
row of Table 1.

\begin{table} 
\centering
\begin{tabular}{|ccc|c|c|}
\hline $a_1$ & $a_2$ & $a_3$ & $\bar{\lambda}$ &  $\hat{\lambda}_M$ \\
\hline
 1&1&1        &.001078    &.001099\\ \hline
 10 & 10 & 10 &.001402    &.001110\\ \hline
 10 & 10 & 20 &.001515    &.001113\\ \hline
 10 & 20 & 20 &.001643   &.001117\\ \hline
 20 & 20 & 20 &.001739    &.001142\\ \hline
 30 & 30 & 30 &.002105   &.001135\\ \hline
\end{tabular}
\caption{ \label{t3} Tho methods for estimating the null occurrence
rate of palindrome sequences are compared when non-random clusters
exist. For each random sequence, three non-random clusters are
inserted with adjustable occurrence rates: $\lambda_i=a_i\lambda_0$,
$1\le i\le 3$. $\lambda_0=.00098$. The first row, with
$a_1=a_2=a_3=1$, means complete random sequence with no hot. }
\end{table}

The validity that the average rate can be a null parameter estimator
is based on the assumption that the number of events from non-random
clusters is much smaller than the total number of events. However,
this assumption may not work for a real DNA sequence. It has been
observed that meaningful sites in the sequence tends to have higher
palindrome rates. The average rate usually overestimates the null
occurrence rates. Here, we design a simulation experiment to check
the robustness of the estimates against  hot spot regions.

For each random sequence, we insert three hot spot segments with
length 1000 base pairs at different positions. The inserted segments
contain  palindromes  which are randomly resampled from the
palindrome bank.  The palindrome bank collects all the DNA
palindromes from BoHV1 sequences. We assigned three occurrence rates
for the three segments as $\lambda_i=a_i\lambda_0$, $1\le i\le 3$
and $\lambda_0=.00098$ is estimated by Markov model for BoHV1
sequence. $a_i$'s are to quantify the intensities of hot spots. The
simulation results for various components of $(\lambda_1, \lambda_2,
\lambda_3)$ based on 500 repeats are presented in Table \ref{t3}.
The estimators based on model calculation increase less than $8\%$
while the estimator based on the average rate almost doubles, when
the occurrence rates in the hot-spot regions  increase to 30 folds.

Overestimating the occurrence rate would increase the threshold
value for testing hypothesis and lead to power loss. The simulation
for power comparisons in Table 4 is designed as that of Table 3.
Table \ref{t4} shows the powers for detecting each of the three hot
spot regions of DNA palindromes.  We applied the PLS scores and BWS
scores with window size 1000 bp to scan the whole genome. The
calculation for threshold values follows Chan and Zhang (2007) on
weighted scan statistics, with modification on the overshoot term,
which is shown in the appendix of this paper. Here, power is defined
as the frequencies of detecting hot spot regions based on 500
replicates. Table \ref{t4} shows that $\hat{\lambda}_M$ can gain
powers more than 50$\%$ over $\bar{\lambda}$, when power is not
saturated.

\begin{table} 
\centering
\begin{tabular}{|c|c|ccc|c|ccc|}
\hline PLS &\multicolumn{8}{c|}{ }\\
\hline \multirow{2}{*}{$(a_1,a_2,a_3)$}
&\multicolumn{4}{c|}{$\overline{\lambda}$}
&\multicolumn{4}{c|}{$\hat{\lambda}_M$}\\\cline{2-9} & Threshold
&\multicolumn{3}{c|}{Power} & Threshold &\multicolumn{3}{c|}{Power}
\\\hline (1,1,1)    &8.9063&0.0000 &0.0000 &0.0000 &9.0061 &0.0000
&0.0000 &0.0000      \\ \hline (7,7,7)    &9.6221&0.2100 &0.2025
&0.2275 &9.0399 &0.2975 &0.2900 &0.2900      \\ \hline (10,10,10)
&9.9477&0.4550 &0.5075 &0.4800 &9.0496 &0.5825 &0.6250 &0.6325
\\ \hline (10,10,20) &10.3013&0.4300 &0.5100 &0.9875 &9.0686 &0.5950
&0.6575 &0.9950      \\ \hline (10,20,20) &10.6435&0.3825 &0.9900
&0.9775 &9.0877 &0.6350 &0.9975 &0.9975      \\ \hline (20,20,20)
&11.0216&0.9675 &0.9850 &0.9850 &9.1014 &0.9900 &0.9925 &0.9975 \\
\hline
\hline BWS &\multicolumn{8}{c|}{ }\\
\hline \multirow{2}{*}{$(a_1,a_2,a_3)$}
&\multicolumn{4}{c|}{$\overline{\lambda}$}
&\multicolumn{4}{c|}{$\hat{\lambda}_M$}\\\cline{2-9} & Threshold
&\multicolumn{3}{c|}{Power} & Threshold &\multicolumn{3}{c|}{Power}
\\\hline (1,1,1)    &114.4505&0.0000 &0.0000 &0.0000 &115.7137
&0.0000 &0.0000 &0.0000      \\ \hline (7,7,7)    &123.2021&0.1950
&0.2425 &0.2625 &115.9571 &0.2700 &0.3250 &0.3200      \\ \hline
(10,10,10) &127.5283&0.5150 &0.5325 &0.5525 &116.0439 &0.6650
&0.6650 &0.6800      \\ \hline (10,10,20) &130.8699 &0.4575 &0.4625
&0.9800 &116.1847 &0.6425 &0.6325 &0.9925 \\ \hline (10,20,20)
&133.7581&0.4100 &0.9850 &0.9775 &116.1572 &0.6125 &1.0000 &0.9975
\\ \hline (20,20,20) &140.2448 &0.9825 &0.9825 &0.9750 &116.3187
&0.9950 &1.0000 &0.9925  \\ \hline
\end{tabular}
\caption{ \label{t4} Powers are compared  for using $\bar{\lambda}$
and $\hat{\lambda}_M$ to estimate the null occurrence rates of DNA
palindromes when hot spot regions are inserted. $\bar{\lambda}$
tends to be too conservative by overestimate the occurrence rates. }
\end{table}

\section{ Discussion}

Average rate is a popular method for estimating the null occurrence
rate of scan statistics. In this paper, we show  that it does not
always work through an example. Average rate can overestimate the
null occurrence rate twice the true number, in the herpesvirus
genome simulation.  We further proposed a model based estimator,
which avoids to directly count the number of events in hot spot
regions. Our method estimates the Markov parameters instead of
estimating the occurrence rate directly.

The hot spot regions have potential to contribute a large portion of
the number of events, especially when the null occurrence rate is
very low. On the other hand, when estimating the transition
probabilities for transition as well as the stationary state
probabilities under the Markov model, the hot spots have little
influence provided their size is much smaller than the total length
of the genomes. This explains why $\hat{\lambda}_M$ is not sensitive
to the hot spot effect. Our study suggests that average rate should
be carefully used for null parameter estimation, especially when the
process involves rare events with  hot spot regions, which are quite
common in epidemiology studies with rare diseases.

\section{ Appendix}

Chan and Zhang (2007) have provided a $p$-value approximation for
the scan statistics of marked Poisson processes. Here, we provide a
more general formula for calculating the overshoot term on various
distribution of $x_i$. Let $N$ be a Poisson process with constant
rate $\lambda_0
>0$ and let random variables
$x_1,\dots,x_n
 \stackrel{\mathrm{iid}}{ \sim} f_{\theta_0}(\cdot)$. Let
 $\lambda_1$ and $\theta_1$ satisfy two conditions : (a)
$\mbox{~}\lambda_1 \phi'(\theta_1)=b$. (b)
$\mbox{~}\log(\lambda_1/\lambda_0)-
(\phi(\theta_1)-\phi(\theta_0))=0$.  Then we have the following
theorem.

 \noindent{\bf Theorem 4} Let $W \rightarrow \infty$ as $w \rightarrow \infty$ such that
 $W-w \rightarrow \infty$. Then
\begin{eqnarray*}\label{A01}
&&\mathrm{P}_0(\max\limits_{0 <s< W}S_{N_w(s)} \geq b)  \approx
1-exp\bigg \{-(W-w) \nu_{\lambda_1,\theta_1}( b- \lambda_0
\mu_0)e^{-I(b)w} (2 \pi w \lambda_1 \phi''(\theta_1))^{-1/2} \bigg \},\\
&&\mbox{~with~~~~} \nu_{\lambda_1,\theta_1} = \frac{1-\mathrm{E}_0
e^{-S_{\tau_+}(\theta_1-\theta_0)}}{(1-e^{-(\theta_1-\theta_0)
})\mathrm{E}_0 S_{\tau_+}}.
\end{eqnarray*}
 \noindent{\bf Proof of Theorem 4}\\
Assume that the process is observed on the set $\{t_j|t_j=j\Delta,
0\le t_j\le W\}$, where $\Delta=o(w)$, then we have the inequality:

\begin{eqnarray*}
\mathrm{P}(\max_{0 \leq j\Delta \leq W}S_{N_w(j\Delta)} \ge b) \leq
\mathrm{P}( \max\limits_{0  \leq s \leq W} S_{N_w(s)} \ge b) \leq
\mathrm{P}(\max_{0 \leq j \Delta \leq W}S_{N_{w+\Delta}(j \Delta)}
\ge b).
\end{eqnarray*}
{ It can be shown that $ \mathrm{P}( \max\limits_{0  \leq s \leq W}
S_{N_w(s)} \ge b)$ converges when $w$ converges to a constant such
that $\lim_{\Delta\rightarrow 0}\mathrm{P}( \max\limits_{1 \leq i
\leq W/\Delta} S_{N_w(i\Delta)} \ge b)= \mathrm{P}( \max\limits_{0
\leq s \leq W} S_{N_w(s)} \ge b)$. In fact, in this study, if we let
$W$ be the total number of DNA base pairs, then $\Delta$ equals 1
instead of converging to 0.

First, we decompose the probability by the last time conditioning
$\tau_b=sup\{j|S_{N_w(j\Delta)}\ge b\}$ used in (Woodroofe, 1979)
\begin{eqnarray*}
&&\mathrm{P}\big(\max_{0 \leq j\Delta \leq W}S_{N_w(j\Delta)} \geq b\big)
=\sum_{0\leq j \leq \lfloor (W-w)/\Delta
\rfloor  }P(\tau_b=j)\\
&=& \sum\limits^{\lfloor (W-w)/\Delta
\rfloor}_{j=0}\mathrm{P}\bigg\{ \max\limits_{j< s \leq
\lfloor(W-w)/\Delta \rfloor} S_{N_w(s)} <b,
S_{N_w(j\Delta)}\geq b\bigg\}\\
&  \approx& \frac{(W-w)}{\Delta} \sum\limits^{\infty}_{k=0}
\mathrm{P}\bigg\{ \max\limits_{0< j \leq \infty} S_{N_w(j\Delta)}
<b, S_{N_w(0)}=b+k\bigg\}.
\end{eqnarray*}
}
 This approximation technique can be found in (Tu and Siegmund,
1999). We applied the new measure $\mathrm{Q}$ introduced in (Chan
and Zhang, 2007), which $\mathrm{Q}$ is defined as that $\mathrm{N}$
is nonuniform poisson with rate $\lambda_1$ on $(0,w]$ and rate
$\lambda_0$ on $(w,W]$; $x_i
\stackrel{\mathrm{ind.}}{\sim}f_{\theta_1}(\cdot) \mbox{~for~} 1\leq
i \leq N(w)$ and $x_i \stackrel{\mathrm{ind.}}{\sim}
f_{\theta_0}(\cdot) \mbox{~for~} N(w) \leq i \leq N(W)$. By (a) and
(b),
\begin{eqnarray*}
&&\frac{dQ}{dP}\{N,x_1,\dots\,x_{N(W)}\}\\
&=& \exp(S_{N_w(0)}(\theta_1-\theta_0)-(\lambda_1-\lambda_0)w).
\end{eqnarray*}
By change of measure, we have
\begin{eqnarray*}
&&\sum\limits^{\infty}_{k=0}\mathrm{P}(\max\limits_{0 < i \leq \infty }S_{N_w(i \Delta)} <b, S_{N_w(0)}=b+k)\\
&=&\sum\limits^{\infty}_{k=0}\mathrm{E}_{\mathrm{Q}}[
\frac{dP}{dQ}\textbf{1}\{\max\limits_{0 <i \leq \infty }S_{N_w(i \Delta)} <b, S_{N_w(0)}=b+k\}]\\
&=&\sum\limits^{\infty}_{k=0}e^{-I(b)w-k(\theta_1-\theta_0)}\mathrm{Q}(\max_{0<i
\leq \infty }S_{N_w(i\Delta)}-S_{N_w(0)} < -k |
S_{N_w{(0)}}=b+k)\mathrm{Q}(S_{N_w(0)}=b+k),
\end{eqnarray*}
where $I(b)=b(\theta_1-\theta_0)/w-(\lambda_1-\lambda_0) $.\\

By local CLT,
\[
\mathrm{Q}(S_{N_w(0)}=b+k) \approx [2\pi w \lambda_1
\phi''{\theta_1}]^{-1/2}.
\]
 Let $\{N^*(t),
x^*_1,\dots,x^*_{N^*(t)}\}$ be independent with $\{N(t),
x_1,\dots,x_{N(t)}\}$ and $N^*(t)$ be a poisson process with rate
$\lambda_1$ and $x^*$ is distributed from $f_{\theta_1}(\cdot)$; let
$w$ and $b$ be large enough such that
\begin{eqnarray*}
\mathrm{Q}(\max_{0<j \leq \infty}S_{N_w(j\Delta)}-S_{N_w(0)} < -k |
S_{N_w(0)}=b+k) \approx \mathrm{P}\bigg\{\min_{0<j \leq
\infty}(-\sum\limits^{N(j\Delta)}_{k=1}x_k +
\sum\limits^{N^*(j\Delta)}_{k=1}x^*_k)>k\bigg \}
\end{eqnarray*}
Let $y_1 =  (-\sum\limits^{N(\Delta)}_{k=1}x_k +
\sum\limits^{N^*(\Delta)}_{k=1}x^*_k)$, and $y_2$, $y_3, ~\cdots$
are iid copies of $y_1$. By (8.13) in Siegmund(1985), we have
\[ \mathrm{P}(\min\limits_{0<n\leq \infty}S_n
>k)=\frac{\mathrm{P}( S_{\tau_+} > k)\mathrm{E}_0y_1}{\mathrm{E}_0S_{\tau_+}},\mbox{~where~}S_n = \sum\limits^n_{i=1} y_i\mbox{~and~}\tau_+ = \inf\{n :
S_n>0\}.
\]
Since
$\sum\limits^\infty_{k=0}e^{-k(\theta_1-\theta_0)}\mathrm{P}(S_{\tau_+}>k)$
can be expressed as
$(1-\mathrm{E}e^{-S_{\tau_+}(\theta_1-\theta_0)})/(1-e^{-(\theta_1-\theta_0)})$,
we have
\[
\sum\limits^{\infty}_{k=0} \mathrm{P}\bigg\{ \max\limits_{0< s \leq
\infty} S_{N_w(s)} <b, S_{N_w(0)}=b+k\bigg\} \approx
v_{\lambda_1,\theta_1}(\mathrm{E} y_1)e^{-I(b)w} (2\pi w \lambda_1
\phi''(\theta_1))^{-1/2}.
\]
Therefore,
\[ \mathrm{P}(\max\limits_{0 <s< W}S_{N_w(s)} \geq b) \approx 1 - \exp \bigg\{ (W-w)v_{\lambda_1,\theta_1}( b- \lambda_0 \mu_0)e^{-I(b)w}
(2 \pi w \lambda_1 \phi''(\theta_1))^{-1/2} \bigg \}.
\]
 By Theorem 1 of (Tu, 2009), the overshoot
$v_{\lambda_1,\theta_1}$ can be calculated when the characteristic
function $\mathrm{E}e^{ity_1}$ is known. Let
$\mathrm{\phi}(t)=\mathrm{E}e^{itx_1}$. We have
\begin{eqnarray*}
\mathrm{E}[\exp\{-it\sum\limits^{N(\Delta)}_{j=1}x_j\}]
=\sum\limits^{\infty}_{k=0}\mathrm{K}^k(-t)\frac{e^{-\lambda_0\Delta}(\lambda_0\Delta)^k}{k!}
=e^{\lambda_0\Delta(\mathrm{\phi}(-t)-1)}
\end{eqnarray*}
and
\begin{eqnarray*}
\mathrm{E}[\exp\{it\sum\limits^{N^*(\Delta)}_{j=1}x^*_j\}]
=\mathrm{E_Q}[\exp\{it\sum\limits^{N(\Delta)}_{j=1}x_j\}]
=\mathrm{E}[\frac{dQ}{dP}\exp\{it\sum\limits^{N(\Delta)}_{j=1}x_j\}]
=e^{\{-\lambda_1\Delta+\lambda_0\Delta\mathrm{\phi}(t-(\theta_1-\theta_0)i)\}}.
\end{eqnarray*}
So $\mathrm{E}e^{ity_1}$ is derived.

\end{document}